\documentclass[twoside,twocolumn,9pt,nofootinbib]{article}
\usepackage[title]{appendix}
\usepackage{amsfonts}
\usepackage{amsthm}
\usepackage{textcomp}
\usepackage{extsizes}
\usepackage[super,sort&compress,comma]{natbib} 
\usepackage[version=3]{mhchem}
\usepackage[left=1.5cm, right=1.5cm, top=1.785cm, bottom=2.0cm]{geometry}
\usepackage{balance}
\usepackage{mathptmx}
\usepackage{sectsty}
\usepackage{graphicx}   
\usepackage{lastpage}
\usepackage[format=plain,justification=justified,singlelinecheck=false,font={stretch=1.125,small,sf},labelfont=bf,labelsep=space]{caption}
\usepackage{float}
\usepackage{fancyhdr}
\usepackage{fnpos}
\usepackage[english]{babel}
\addto{\captionsenglish}{%
  \renewcommand{\refname}{Notes and references}
}
\usepackage[normalem]{ulem}
\usepackage{upgreek}  
\usepackage{array}
\usepackage{droidsans}
\usepackage{charter}
\usepackage[T1]{fontenc}
\usepackage[usenames,dvipsnames]{xcolor}
\usepackage{setspace}
\usepackage[compact]{titlesec}
\usepackage[colorlinks=true,
            linkcolor=black,
            urlcolor=black,
            citecolor=black,
            bookmarks=false,
            pdfstartview=FitH]{hyperref}
                  
\usepackage{epstopdf}%This line makes .eps figures into .pdf - please comment out if not required.

\definecolor{cream}{RGB}{222,217,201}

\begin{document}

\pagestyle{fancy}
\thispagestyle{plain}
\fancypagestyle{plain}{
%%%HEADER%%%
\renewcommand{\headrulewidth}{0pt}
}
%%%END OF HEADER%%%

%%%PAGE SETUP - Please do not change any commands within this section%%%
\makeFNbottom
\makeatletter
\renewcommand\LARGE{\@setfontsize\LARGE{15pt}{17}}
\renewcommand\Large{\@setfontsize\Large{12pt}{14}}
\renewcommand\large{\@setfontsize\large{10pt}{12}}
\renewcommand\footnotesize{\@setfontsize\footnotesize{7pt}{10}}
\makeatother

\renewcommand{\thefootnote}{\fnsymbol{footnote}}
\renewcommand\footnoterule{\vspace*{1pt}% 
\color{cream}\hrule width 3.5in height 0.4pt \color{black}\vspace*{5pt}} 
\setcounter{secnumdepth}{5}

\makeatletter 
\renewcommand\@biblabel[1]{#1}            
\renewcommand\@makefntext[1]% 
{\noindent\makebox[0pt][r]{\@thefnmark\,}#1}
\makeatother 
\renewcommand{\figurename}{\small{Fig.}~}
\sectionfont{\sffamily\Large}
\subsectionfont{\normalsize}
\subsubsectionfont{\bf}
\setstretch{1.125} %In particular, please do not alter this line.
\setlength{\skip\footins}{0.8cm}
\setlength{\footnotesep}{0.25cm}
\setlength{\jot}{10pt}
\titlespacing*{\section}{0pt}{4pt}{4pt}
\titlespacing*{\subsection}{0pt}{15pt}{1pt}
%%%END OF PAGE SETUP%%%

%%%FOOTER%%%
\fancyfoot{}
\fancyfoot[LO,RE]{\vspace{-7.1pt}\includegraphics[height=9pt]{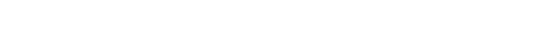}}
\fancyfoot[CO]{\vspace{-7.1pt}\hspace{13.2cm}\includegraphics{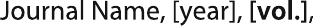}}
\fancyfoot[CE]{\vspace{-7.2pt}\hspace{-14.2cm}\includegraphics{head_foot/RF}}
\fancyfoot[RO]{\footnotesize{\sffamily{1--\pageref{LastPage} ~\textbar  \hspace{2pt}\thepage}}}
\fancyfoot[LE]{\footnotesize{\sffamily{\thepage~\textbar\hspace{3.45cm} 1--\pageref{LastPage}}}}
\fancyhead{}
\renewcommand{\headrulewidth}{0pt} 
\renewcommand{\footrulewidth}{0pt}
\setlength{\arrayrulewidth}{1pt}
\setlength{\columnsep}{6.5mm}
\setlength\bibsep{1pt}
%%%END OF FOOTER%%%

%%%FIGURE SETUP - please do not change any commands within this section%%%
\makeatletter 
\newlength{\figrulesep} 
\setlength{\figrulesep}{0.5\textfloatsep} 

\newcommand{\topfigrule}{\vspace*{-1pt}% 
\noindent{\color{cream}\rule[-\figrulesep]{\columnwidth}{1.5pt}} }

\newcommand{\botfigrule}{\vspace*{-2pt}% 
\noindent{\color{cream}\rule[\figrulesep]{\columnwidth}{1.5pt}} }

\newcommand{\dblfigrule}{\vspace*{-1pt}% 
\noindent{\color{cream}\rule[-\figrulesep]{\textwidth}{1.5pt}} }

\makeatother
%%%END OF FIGURE SETUP%%%

%%%TITLE, AUTHORS AND ABSTRACT%%%
\twocolumn[
  \begin{@twocolumnfalse}
{\hfill\raisebox{0pt}[0pt][0pt]{\includegraphics[height=25pt]{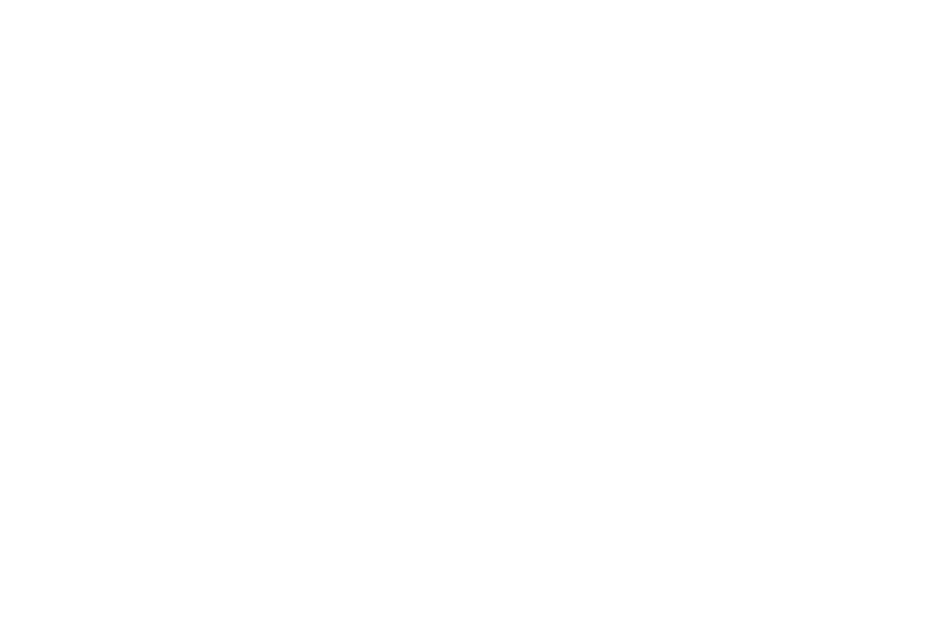}}\\[1ex]
}\par
\vspace{1em}
\sffamily
\begin{tabular}{m{4.5cm} p{13.5cm} }

 & \noindent\LARGE{\textbf{Crack patterns of drying dense bacterial suspensions$^\dag$}} \\%Article title goes here instead of the text "This is the title"
\vspace{0.3cm} & \vspace{0.3cm} \\

 & \noindent\large{Xiaolei Ma,$^{\ast}$\textit{$^{a}$} Zhengyang Liu,\textit{$^{a}$} Wei Zeng,\textit{$^{ab}$} Tianyi Lin,\textit{$^{a}$} Xin Tian,\textit{$^{c}$} and Xiang Cheng$^{\ast}$\textit{$^{a}$}} \\%Author names go here instead of "Full name", etc.

{\includegraphics[height=60pt]{head_foot/RSC_LOGO_CMYK}}  & \noindent\normalsize{Drying of bacterial suspensions is frequently encountered in a plethora of natural and engineering processes. However, the evaporation-driven mechanical instabilities of dense consolidating bacterial suspensions have not been explored heretofore. Here, we report the formation of two different crack patterns of drying suspensions of \textit{Escherichia coli} (\textit{E. coli}) with distinct motile behaviors.  Circular cracks are observed for wild-type \textit{E. coli} with active swimming, whereas spiral-like cracks form for immotile bacteria. Using the elastic fracture mechanics and the poroelastic theory, we show that the formation of the circular cracks is determined by the tensile nature of the radial drying stress once the cracks are initiated by the local order structure of bacteria due to their collective swimming. Our study demonstrates the link between the microscopic swimming behaviors of individual bacteria and the mechanical instabilities and macroscopic pattern formation of drying bacterial films. The results shed light on the dynamics of active matter in a drying process and provide useful information for understanding various biological processes associated with drying bacterial suspensions.}%The abstrast goes here instead of the text "The abstract should be..."

\end{tabular}

\end{@twocolumnfalse} \vspace{0.6cm}

  ]
%%%END OF TITLE, AUTHORS AND ABSTRACT%%%

%%%FONT SETUP - please do not change any commands within this section
\renewcommand*\rmdefault{bch}\normalfont\upshape
\rmfamily
\section*{}
\vspace{-1cm}

%%%FOOTNOTES%%%

\footnotetext{\textit{$^{a}$~Department of Chemical Engineering and Materials Science, University of Minnesota, Minneapolis, MN 55455, USA; E-mail: iamxlma@gmail.com; xcheng@umn.edu}}
\footnotetext{\textit{$^{b}$~College of Life Science and Technology, Guangxi University, Nanning 530004, Guangxi, China}}
\footnotetext{\textit{$^{c}$~Department of Physics \& Astronomy, University of Wyoming, Laramie, WY 82071, USA}}

%Please use \dag to cite the ESI in the main text of the article.
%If you article does not have ESI please remove the the \dag symbol from the title and the footnotetext below.
\footnotetext{\dag~Electronic Supplementary Information (ESI) available. See DOI: 10.1039/cXsm00000x/}
%additional addresses can be cited as above using the lower-case letters, c, d, e... If all authors are from the same address, no letter is required

%\footnotetext{\ddag~Additional footnotes to the title and authors can be included \textit{e.g.}\ `Present address:' or `These authors contributed equally to this work' as above using the symbols: \ddag, \textsection, and \P. Please place the appropriate symbol next to the author's name and include a \texttt{\textbackslash footnotetext} entry in the the correct place in the list.}

%%%END OF FOOTNOTES%%%

%%%MAIN TEXT%%%%
\section{Introduction}
Active matter is a class of nonequilibrium systems consisting of autonomous units that convert local internal or ambient free energy into mechanical motions. A large number of biological and physical systems including suspensions of self-propelled cytoskeleton\cite{Schaller2010motility,Duclos2020microtubules}, swarms of bacteria\cite{Guo2018ShearBanding,peng2021imaging,liu2021density} and clusters of synthetic active colloids\cite{Palacci2013colloids,Bricard2013colloids,Karani2019colloids} can be categorized as active matter, which exhibit fascinating statistical and mechanical properties that draw tremendous attention in recent years due to their fundamental and technical interests \cite{Ramaswamy2010Review,marchetti2013hydrodynamics,gompper20202020}. While extensive studies have been conducted in understanding the emergent collective dynamics of active matter in fluid states \cite{Saintillan2015bacteria,Alert2022turbulence}, it is still unclear whether and how local activity affects the macroscopic mechanical properties of consolidating active matter during a drying process.  

Drying of bacterial suspensions as a premier model of active matter plays a crucial role in many biological, environmental and industrial processes and influences diverse phenomena ranging from biofilm formation \cite{epstein2012liquid}, spreading of disease \cite{hall2004bacterial} and food hygiene \cite{kumar1998significance} to interbacterial competition for survival \cite{yanni2017life}, coating and self-assembly \cite{nellimoottil2007evaporation,sempels2013auto,andac2019active}. The active swimming of bacteria can profoundly modify the complex interplays between solid, liquid and vapor phases as a suspension passes from a fluid to a solid state during drying, giving rise to the unusual growth dynamics and morphologies of ``coffee rings'' in dried deposits \cite{nellimoottil2007evaporation,kasyap2014bacterial,sempels2013auto,andac2019active,agrawal2020dynamics}. However, these existing studies are all limited to the dilute limit of bacterial suspensions, where bacteria are deposited near the edge of drying drops. The mechanical instabilities of thick consolidating bacterial films formed by drying dense bacterial suspensions remain elusive, despite that such instabilities and the resulting crack patterns have been extensively investigated in counterpart passive particle systems \cite{goehring2015desiccation,allain1995regular,dufresne2003flow,jing2012formation,dugyala2016role,lama2021synergy,domokos2020plato,
ma2019universal,routh2013drying,zang2019evaporation}.

Here, we explored the effect of bacterial swimming on the mechanical instabilities and desiccation crack patterns of dense consolidating bacterial suspensions. We used a wild-type strain of \textit{Escherichia coli} (\textit {E. coli}) as our model bacteria, which display the classic run-and-tumble swimming in water \cite{berg2008coli}. As a control, we also examined a mutant strain of \textit {E. coli} that show only tumbling. Below, we shall refer to the wild-type \textit {E. coli} as swimmers and the mutant \textit {E. coli} as tumblers. While spiral-like cracks were found in the dried deposits of tumblers and dead swimmers, we observed circular cracks in the dried deposits of swimmers. Using the elastic fracture mechanics and the poroelastic theory, we showed that the circular cracks form due to the tensile nature of the radial drying stress once the cracks are initiated by the local order structure induced by the collective swimming of wild-type \textit{E. coli}. In contrast, the spiral-like cracks arise from a dynamic interplay between cracking and delamination of the drying films of immotile bacteria. Our study unambiguously demonstrates the crucial effect of bacterial swimming on the mechanical instabilities of consolidating bacterial films and illustrates the unique features of active matter in infamously complicated drying processes. Our results are also helpful for deciphering different desiccation crack patterns of dried bacterial films encountered in natural and engineering processes.     

\begin{figure}[!]
\begin{center}
\includegraphics[width=3.4 in]{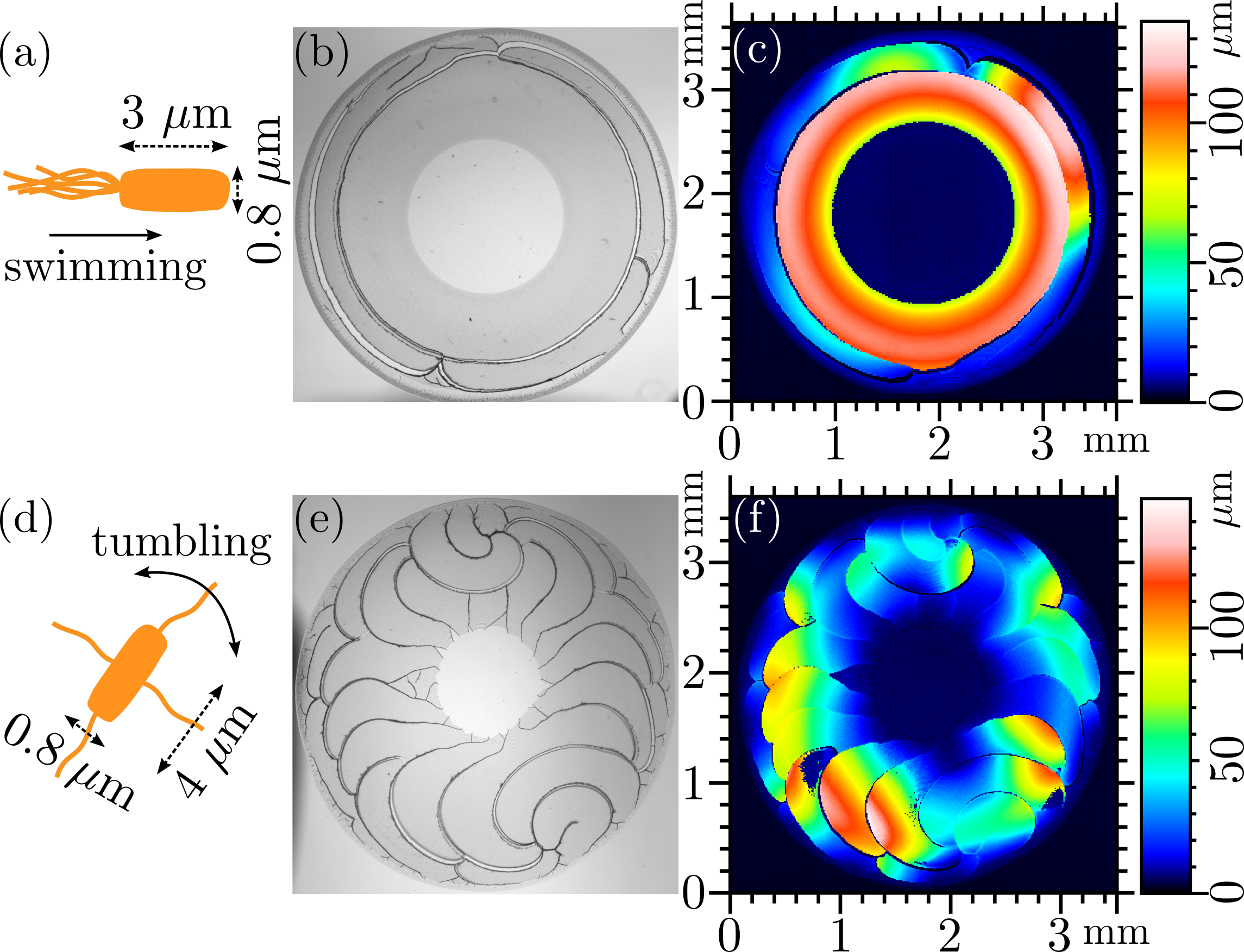}
\caption[]{Schematics of a wild-type \textit{E. coli} with run-and-tumble motion (a) and a mutant \textit{E. coli} with tumbling motion only (d). Bright-field images of the dried deposits of a drop of the wild-type swimmer suspension (b) and a drop of the mutant tumbler suspension (e). Drop volume $V_i=$ 2.5 $\mu$L. Bacterial volume fraction $\phi_i=$ 14 \%. The corresponding height profiles of the deposits are shown in (c) (swimmer) and (f) (tumbler).
\label{schematic_setup}} 
\end{center} 
\end{figure}

\section{Experiment}

We used two different \textit{E. coli} strains with distinct swimming behaviors in our experiments, i.e., a wild-type strain of swimmers (BW25113) and a mutant strain of tumblers (RP1616). The two strains share a similar body geometry, which has the average length of 3-4 $\mu$m and the average width of $0.8$ $\mu$m (Figs.~\ref{schematic_setup}a and d). The culturing protocols of the two strains are detailed elsewhere\cite{liu2021density,peng2021imaging}. In addition, we have also studied the drying behaviors of suspensions of dead swimmers. The dead bacteria were obtained from suspensions of active wild-type swimmers sitting in sealed micro-centrifuge tubes for at least four days. The bacteria were confirmed to be immotile from direct optical microscopy. To avoid any potential complication due to the change of buffer quality over the long waiting, we washed the dead bacteria and resuspended them in DI water at the targeted concentration before each experiment.

In a typical experiment, we prepared a suspension of bacteria with an initial volume fraction $\phi_{i}$ of 10-20\%. A microscope glass slide cleaned with DI water and dried by a blowgun was used as the substrate, which was hydrophilic with a water contact angle $\approx $ 25$^\circ$. A drop of the suspension with an initial volume $V_{i}$ of 2-3 $\mu$L was gently deposited onto the substrate for drying. A bright-field inverted microscope was used to image the drying process at a frame rate of 1-20 fps. All experiments were performed at a room temperature of 20 $\pm$ 2 $^\circ$C with a relative humidity of $RH = 30 \pm 4\%$. We used a scanning electron microscope (SEM) to image the microstructures of dried bacterial deposits, and an optical non-contact profilometer to measure the temporal evolution of the height profiles of drying bacterial films. 

\begin{figure}[!]
\begin{center}
\includegraphics[width=3.5 in]{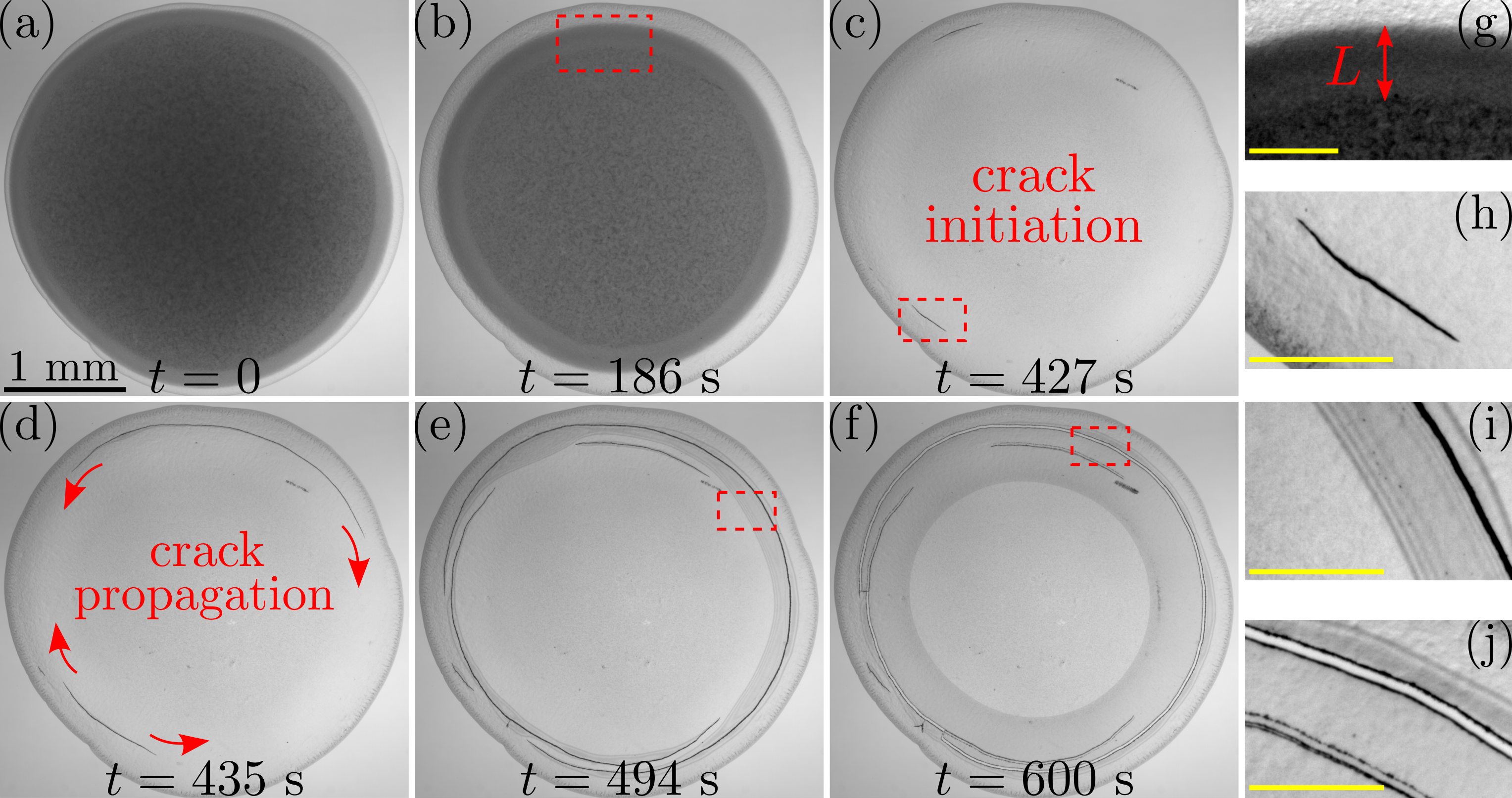}
\caption[]{Snapshots (a)-(f) show the characteristic stages during the formation of circular cracks by drying a drop of wild-type swimmers with $V_{i}=$ 2.5 $\mu$L and $\phi_{i}=14\%$. The drying process lasted about 10 min before the film completely dried. Image (g) shows the zoom-in view of the red boxed area in (b), where $L$ indicates the width of the compaction front. Image (h) shows the zoom-in view of the firstly initiated crack in the boxed region in (c). Images (i) and (j) show the zoom-in views of regions near cracks in the boxes in (e) and (f), respectively. The scale bars in (g)-(j) are 250 $\mu$m. Video S1 in ESI$^\dag$ shows the complete drying process corresponding to (a)-(f).
\label{Fig:cracks_swimmers}
} 
\end{center} 
\end{figure}
\section{Results}
\subsection{Circular cracks of wild-type swimming bacteria}
We observed circular cracks in the dried deposits of suspensions of wild-type swimmers (Figs.\ \ref{schematic_setup}b and c). Figures~\ref{Fig:cracks_swimmers}a-f show the snapshots of different stages during the formation of circular cracks by drying a drop of a swimmer suspension with $V_i=$ 2.5 $\mu$L and $\phi_i=13\%$. After the drop was deposited on the substrate (Fig.~\ref{Fig:cracks_swimmers}a), evaporation initially occurred predominantly near the pinned contact line of the drop, driving the formation of a compaction front where the concentration of bacteria increased drastically from $\phi_i$ in the bulk fluid to that close to the random close packing in the consolidating film $\phi$. The compaction front displayed an approximately constant length $L$ (Fig.\ \ref{Fig:cracks_swimmers}g), and continuously moved toward the center of the drop over a time interval of 414 s (Fig.\ \ref{Fig:cracks_swimmers}b and Video S1 in ESI$^\dag$). After the passing of the compaction front, the bacterial film was wet and gel-like and continued to undergo evaporation along the top surface of the film, leading to the accumulation of stress. Once the critical material strength was reached around 427 s, cracks were initiated near the edge of the film to release the excess stress \cite{griffith1921vi} (Figs.\ \ref{Fig:cracks_swimmers}c and \ref{Fig:cracks_swimmers}h), which then propagated along a circular path (Figs.\ \ref{Fig:cracks_swimmers}d-\ref{Fig:cracks_swimmers}e) over a short interval of about 73 s in a stick-slip fashion and eventually formed the circular cracks (Fig.\ \ref{Fig:cracks_swimmers}f). Accompanying the crack propagation, the film also delaminated radially toward the drop's center as indicated by the interference fringes in Figs.\ \ref{Fig:cracks_swimmers}i-\ref{Fig:cracks_swimmers}j. %The extent of film delamination in the final dried deposit can be visualized from the height profile of the deposit in Fig.\ \ref{schematic_setup}c. 

\begin{figure*}[!]
\begin{center}
\includegraphics[width=6.3 in]{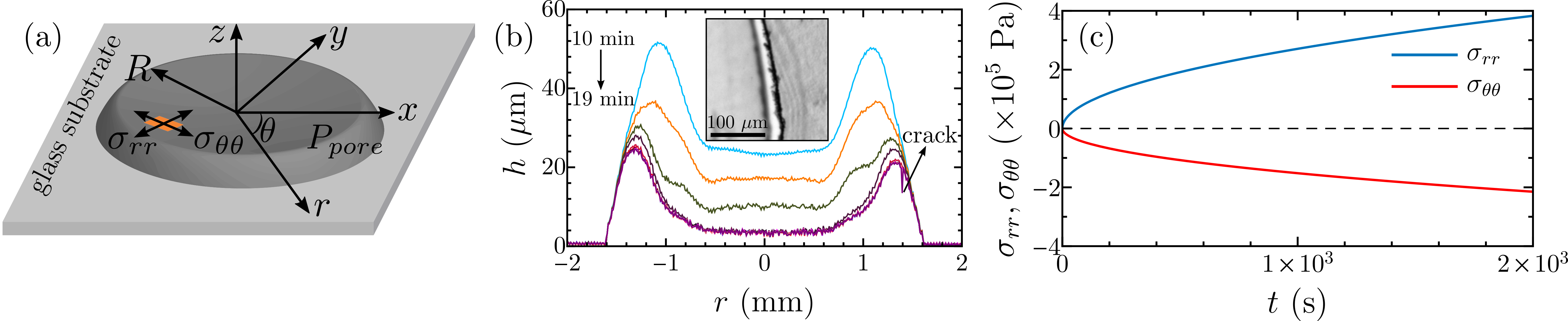}
\caption[]{(a) Illustration of the geometry of a drying drop. (b) The temporal variation of the height profile $h$ of a drying wild-type swimmer drop ($V_i$ = 2.5 $\mu$L, $\phi_i$ = 14\%) along its diameter from $t$ = 10 min to $t=19$ min during which a circular crack forms. The inset shows the segment of the circular crack at $r\approx 1.4$ mm near the scanned region. (c) The in-plane stress components $\sigma_{rr}$ and $\sigma_{\theta \theta}$ from Eqs.\ \ref{radialstress_swimmer} and \ref{azimuthalstress_swimmer} as a function of $t$ at $r^\star=0.75 R$, where $R = 2$ mm is the radius of the drying drop. 
\label{dryinggeometry}
} 
\end{center} 
\end{figure*}

To interpret the circular cracks of wild-type swimmers, we calculate the stress distributions in the consolidating bacterial film. Figure\ \ref{dryinggeometry}a illustrates the geometry of a consolidating film in a cylindrical coordinate. Drying stress was accumulated in the film behind the compaction front in response to the continuously decreasing local liquid pore pressure $P_{pore}$ over time. As evaporation proceeded, the film became flat and thin and evaporation predominantly occurred at the top surface of the film in the late stage of drying. Therefore, $P_{pore}$ satisfies a one-dimensional diffusion equation with $z=0$ defined at the top surface of the film \cite{biot1941general,giorgiutti2018drying}:     
\begin{equation}
\frac{\partial P_{pore}}{\partial t}=\frac{\kappa E}{\eta}\frac{\partial^2P_{pore} }{\partial z^2},
\label{porepressure}
\end{equation}
where $\eta \approx 10^{-3} \mathrm{\ Pa} \cdot \mathrm{s}$ is the dynamic viscosity of water and $E$ is the Young's modulus of the dehydrated film on glass substrate. We estimated $E \sim$ 100 MPa based on our direct measurement using atomic force microscopy, which is consistent with the reported modulus of an isolated dehydrated bacterium\cite{yao1999thickness}. Although the Young's moduli of isolated bacteria and bacterial films are of the same order of magnitude at the microscopic scales, the latter is generally larger depending on substrate stiffness, humidity, bacterial type and even the fitting model \cite{kundukad2016mechanical,even2017recent,chen2012bacterial,abe2011elasticity}. Here, $\kappa$ is the permeability of the film, which is given by the Carman–Kozeny relation $\kappa=\frac{1}{45} \frac{\left(1-\phi\right)^{3}}{\phi^{2}} a^{2} \sim  10^{-15}$ m$^2$, $\phi\approx 0.7$ is the packing fraction of bacteria (Appendix C) and $a \approx 1$ $\mu$m is the characteristic size of bacteria. The initial condition is $P_{pore}(z, 0)= P_{atm}$, where $P_{atm} = 10^5$ Pa is the atmospheric pressure. The boundary conditions is $\partial P_{pore}/\partial z|_{z=0}=-\eta V_{E}/\kappa $, where $V_{E}$ is the steady surface evaporation rate. The evaporation rate is expressed as $V_{E}=\frac{D_{w}}{R}\frac{n_{wsat}}{n_{w}}A(\theta)(1-RH)$ \cite{giorgiutti2014elapsed}, in which $D_w= 25 \times 10^{-6}$ m$^2$ s is the diffusion coefficient of water into air at the room temperature, $R \approx 2$ mm is the drop radius pinned at the contact line, $n_{wsat}=0.02\ \mathrm{kg/m^3}$ is the water density in the vapor at the air-water interface, $n_w= 1 \times 10^{3}\ \mathrm{kg/m^3}$ is the water density in the bulk liquid, $A(\theta)\approx 1.3$ for a contact angle $\theta \approx 25^\circ$, and the relative humidity $RH \approx 30\%$. Therefore, $V_{E} \sim 5 \times 10^{-7}$ m/s. Solving Eq.~\ref{porepressure}, we have $P_{pore}$ given by \cite{style2011mud}:
\begin{equation}
\label{pore_pressure_expression}
P_{pore}(z, t)  =  P_{atm}-P^\star(z,t),
\end{equation}
where  
\begin{equation}
\label{pore_pressure_difference}
P^\star(z,t) = \frac{2 \eta V_E}{\kappa} \left[ \sqrt{\frac{ct}{\pi}} \mathrm{e}^{\frac{-z^{2}}{ 4 ct}}+\frac{z}{2} \operatorname{erfc}\left( \frac{-z}{\sqrt{4ct}}\right) \right]
\end{equation}
represents the deviation of the pore pressure from the atmospheric pressure. Here, $c \equiv E \kappa /\eta$ is a constant.   

\begin{figure*}[!]
\begin{center}
\includegraphics[width=7.0 in]{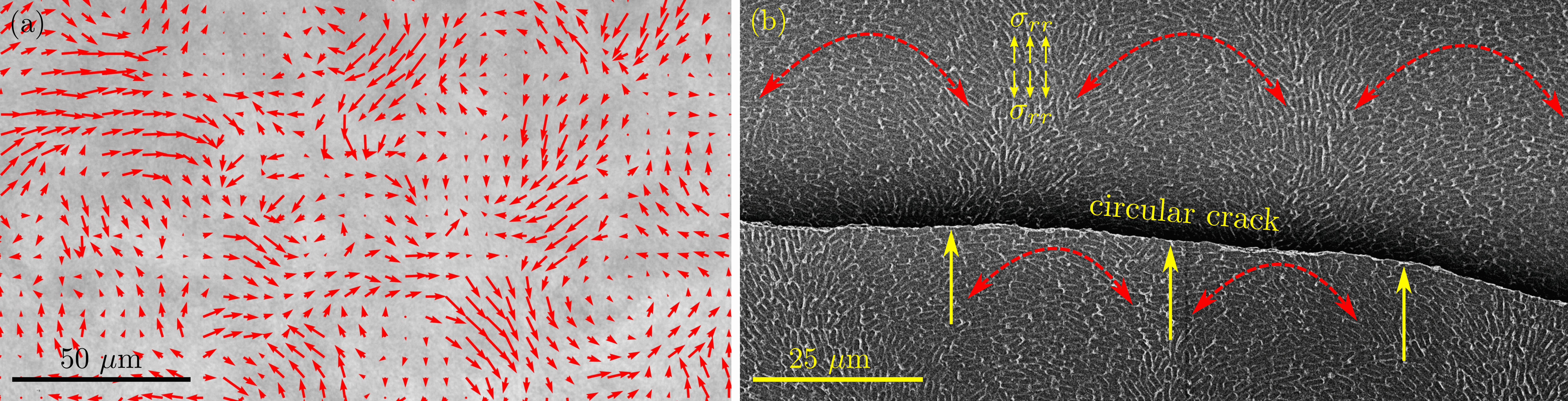}
\caption[]{(a) Snapshot of the 2D flow field of a swimmer suspension near the compaction front during drying. (b) SEM image of the local order structure (red arrows) near a segment of a circular crack (yellow arrows) in a dried swimmer deposit. Video S2 in ESI$^\dag$ shows the 2D flow field of the drying swimmer suspension corresponding to (a).
\label{Fig:SEM_swimmers}
} 
\end{center} 
\end{figure*}

During consolidation, a film is constricted by the substrate, thus the out-of-plane strain $\epsilon_{zz}$ is significantly larger than the in-plane strain $\epsilon_{rr}+\epsilon_{\theta\theta}$ driven by shrinkage. Figure\ \ref{dryinggeometry}b shows the temporal variation of the height profile $h(r)$ along a diameter of a consolidating film from 10 min to 19 min, during which a circular crack forms as indicated by the sharp drop of $h$ at $r\approx 1.4$ mm. The film thickness $h$ decreases significantly during this period, whereas the in-plane contraction is negligible, suggesting that $\epsilon_{zz}\gg \epsilon_{rr}+\epsilon_{\theta\theta}$. Furthermore, since the top surface of the film is traction-free, the deformation of the thin film is predominantly driven by the in-plane stress, suggesting $\sigma_{zz}\ll \sigma_{rr}+\sigma_{\theta\theta}$ \cite{giorgiutti2014elapsed,giorgiutti2018drying}. The same strain and stress conditions have also been applied in modeling the drying of colloidal suspensions on rigid glass substrates\cite{lama2021synergy}.

The in-plane stress components $\sigma_{rr}$ and $\sigma_{\theta\theta}$ responsible for crack formation are correlated by the equilibrium condition of the stress field in the cylindrical coordinate:
\begin{equation}
\frac{\partial \sigma_{r r}}{\partial r}+\frac{1}{r}\left(\sigma_{r r}-\sigma_{\theta \theta}\right)=0.
\label{radial_azimuthal_stress}
\end{equation}
Using the Biot constitutive relation of homogeneous and isotropic solids \cite{biot1941general} under the approximations $\epsilon_{zz}\gg \epsilon_{rr}+\epsilon_{\theta\theta}$ and $\sigma_{zz} \ll \sigma_{rr}+\sigma_{\theta\theta}$, we have (Eq.~\ref{in-plane_stress} in Appendix A): 
\begin{equation}
    \sigma_{rr} + \sigma_{\theta\theta} = \frac{2P^\star(1-2\nu)}{1-\nu},
    \label{Biot_constiutive}
\end{equation}
where $\nu$ is Poisson's ratio of the bacterial film. Given the boundary condition $\sigma_{rr}|_{r = R} = P^\star$, $\sigma_{rr}$ and $\sigma_{\theta\theta}$ can be analytically solved with Eqs.\ \ref{radial_azimuthal_stress} and \ref{Biot_constiutive} (Eqs.~\ref{sol:sigmarr} and \ref{sol:sigmatheta} in Appendix A): 
\begin{eqnarray}
\label{radialstress_swimmer}
\sigma_{rr}=\left\langle P^\star \right\rangle \frac{r^2(2\nu-1)-R^2\nu}{r^2(\nu-1)},\\
\label{azimuthalstress_swimmer}
\sigma_{\theta\theta} = \left\langle P^\star \right\rangle \frac{r^2(2\nu-1)+R^2\nu}{r^2(\nu-1)},
\end{eqnarray}
where $\left\langle P^\star \right\rangle$ the film-thickness-averaged liquid pressure (Eq.~\ref{eq:p_pore_average} in Appendix A).  

It has been shown that biofilms including those of \textit{E. coli} behave mechanically similar to polymeric materials, which have Poisson's ratios $\nu$ ranging between 0.4 and 0.5. The range has been adopted in many previous experimental, numerical and theoretical studies on biofilms\cite{robinson2014encyclopedia,laspidou2007variation,gan2016microrheology,kandemir2018mechanical,pagnout2019pleiotropic,rmaile2013microbial,kundukad2016mechanical,yan2019mechanical}. Here, we took the lower bound of $\nu= 0.4$ to accommodate the condition that the bacteria film was still undergoing drying at the point of cracking. Choosing any value between 0.4 and 0.5 would not qualitatively change our results. With $R \approx 2$ mm and $h \approx 20$ $\mu$m (the thickness of the film at which the crack is generated, see Fig.\ \ref{dryinggeometry}b), Eqs.\ \ref{radialstress_swimmer} and \ref{azimuthalstress_swimmer} predict the spatiotemporal stress distributions within the drying film. 

Figure\ \ref{dryinggeometry}c shows $\sigma_{rr}(r^\star,t)$ and $\sigma_{\theta\theta}(r^\star,t)$ at a fixed location $r^\star = 0.75R$ where the cracks formed in the experiment. Note that $\sigma_{rr}$ is tensile ($\sigma_{rr}>0$) and reaches $\sigma_c \sim$ 10$^5$ Pa when $t > 200$ s, suggesting the formation of the cracks in the circumferential direction at large times. Here, $\sigma_c$ is the critical stress for cracking, which can be estimated based on the the critical film thickness at which cracks form (Appendix B). In comparison, $\sigma_{\theta \theta}$ is compressive ($\sigma_{\theta \theta} < 0$), precluding the formation of cracks in the radial direction that are commonly observed in the drying films of colloidal suspensions \cite{giorgiutti2014elapsed,bourrianne2021crack} based on the principle of fracture mechanics. Note that for drying colloidal suspensions with a typical Poisson's ratio of $\nu = 0.2$,~\cite{leang2017crack,style2011crust,badar2022moving,di2012rheological} $\sigma_{\theta\theta} > 0$ is tensile, which allows for the formation of radial cracks.

In addition to changing the mechanical properties of drying bacterial films, the active swimming of bacteria also modifies the microscopic structure of the consolidating bacterial films, which further facilitates the formation of circular cracks. Figure\ \ref{Fig:SEM_swimmers}a shows the two-dimensional (2D) flow field of a suspension of wild-type swimmers near the compaction front during drying, which exhibits the characteristic swarming vortices induced by bacterial collective swimming\cite{peng2021imaging,liu2021density}. Such a coherent dynamic structure was preserved throughout the drying course and gave rise to the local order packing of bacteria in the dried deposit as indicated by the red arrows in Fig.\ \ref{Fig:SEM_swimmers}b. The circular crack marked by the yellow arrows propagated through this local order structure. 

It has already been shown that drying cracks prefer to propagate along the order direction of underlying microstructures due to the least resistance to release stress during crack propagation\cite{dugyala2016role}. In the simple case of drying suspensions of colloidal ellipsoids where the ellipsoids align parallel to the contact line of drying films, the effect leads to the formation of circular cracks along the circumference of dried annular deposits\cite{dugyala2016role}. In a drying bacterial film, although the swarming vortices do not align along the circumference of the drying film, the order structure should still act locally as a nucleation point, promoting the formation of the initial crack at small bacterial scales. As swarming vortices are mesoscopic on the order of a few tens of microns, the structure of the drying film at large scales is still isotropic. Thus, upon nucleation and after the size of the initial crack reaches the scale of a single swarming vortex, the drying film can be treated as a homogeneous medium with isotropic mechanical properties for the propagation of cracks at large scales. The radial tensile stress calculated in our simple homogeneous model should then be valid, which directs the propagation of cracks and eventually leads to the formation of circular cracks in the dried deposit.

\begin{figure}[!]
\begin{center}
\includegraphics[width=3.2 in]{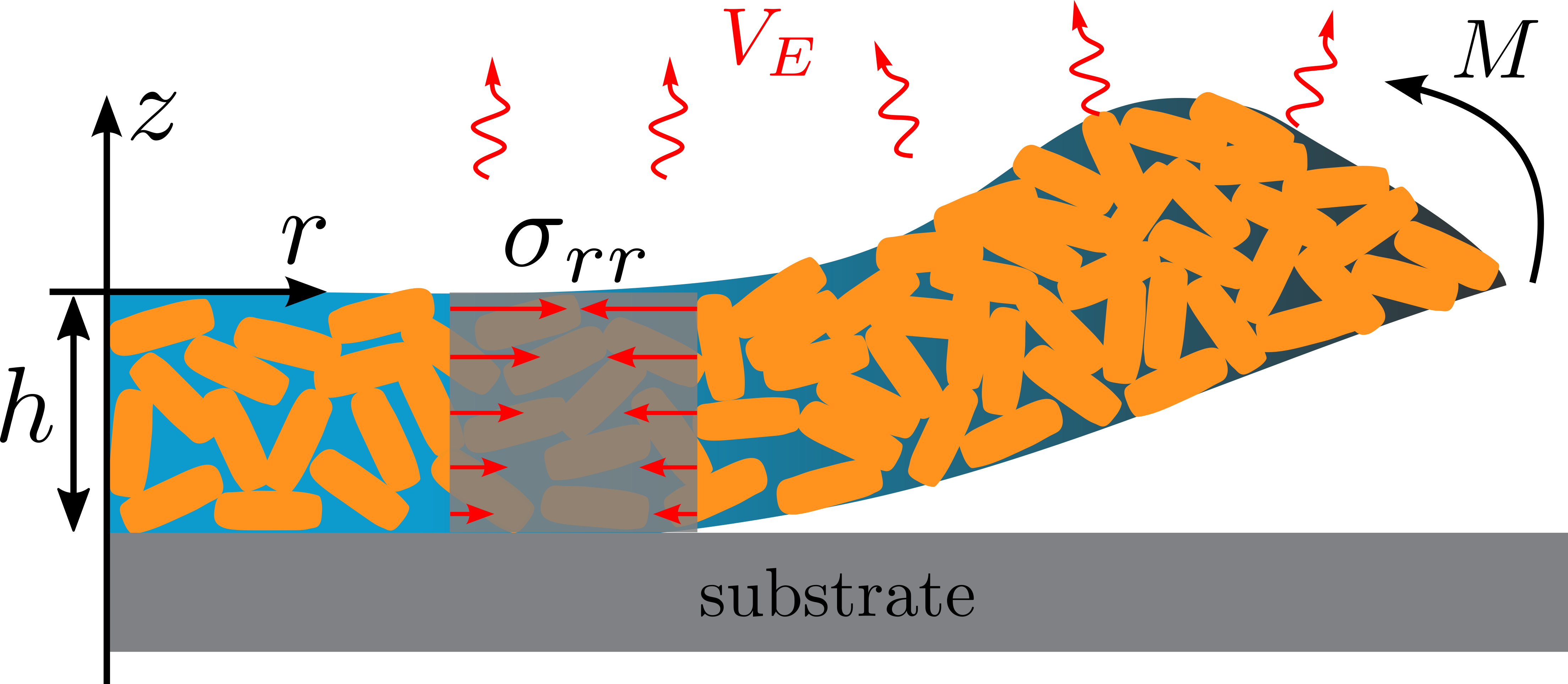}
\caption[]{Schematic illustrating the delamination of a consolidating swimmer film in the radial direction driven by the gradient of the tensile stress $\sigma_{rr}$ across the film thickness.
\label{Fig:delamination_sketch}
} 
\end{center} 
\end{figure}

Finally, Figs.\ \ref{Fig:cracks_swimmers}i-j show that along with the propagation of the circular cracks, the film also delaminated toward the center of the drying drop. Owing to the strong humidity gradient across the film thickness during drying \cite{jagla2002stable}, the in-plane tensile stress $\sigma_{rr}$ is localized near the top surface of the film, which gives rise to a stress gradient across the film as illustrated in Fig.\ \ref{Fig:delamination_sketch}. Such a stress gradient creates a bending moment $M$, promoting the film delamination from the edge of the drop to the center of the drop once the accumulated stress is beyond the critical stress for delamination \cite{jagla2002stable,style2011mud,neda2002spiral}. We estimated both the critical stress for cracking $\sigma_c$ and the critical stress for delamination $\sigma_d$ (Appendix B), which are of the same order of magnitude of $\sim 10^5$ Pa. Hence, immediately following the formation of the circular cracks, the film delaminated from the new crack surface toward the center of the drying drop. Figure\ \ref{schematic_setup}c shows the height profile of a completely-dried swimmer film, which quantitatively illustrates  the extent of film delamination. The part of the film close to the inner edge has a height over 90 $\mu$m above the substrate (the red color), while the thickness of the bacterial film near the pinned contact line (the blue color), as well as in the final stage of the drying process before delamination (Fig.~\ref{dryinggeometry}b), is less than 30 $\mu$m.

\subsection{Spiral-like cracks of immotile bacteria}
\begin{figure}[t]
\begin{center}
\includegraphics[width=3.3 in]{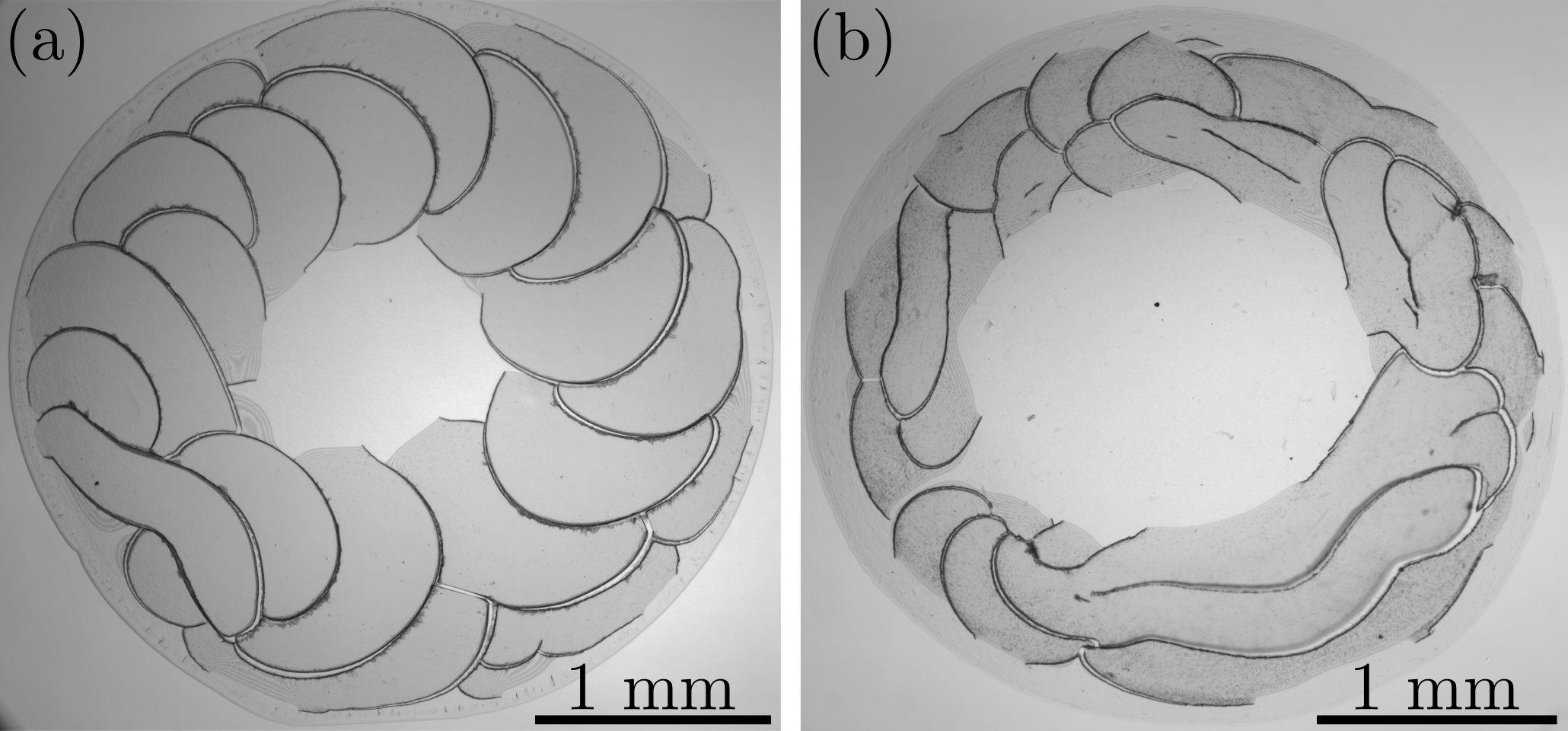}
\caption[]{Bright-field images of the dried deposits of drops of mutant tumblers (a) and dead wild-type swimmers (b) with $V_{i}=$ 2.5 $\mu$L and $\phi_{i}=14\%$. 
\label{Fig:spiral_cracks_dead_swimmer}
} 
\end{center} 
\end{figure}

To verify the role of active swimming in the formation of circular cracks, we further examined the crack patterns of mutant tumblers as well as dead swimmers. Neither of the two types of bacteria showed the collective swimming in drying drops. In contrast to the circular cracks of wild-type swimmers, we observed spiral-like cracks in the dried deposits of these two types of immotile bacteria as shown in Figs.\ \ref{schematic_setup}d-f and \ref{Fig:spiral_cracks_dead_swimmer}.

\begin{figure}[t]
\begin{center}
\includegraphics[width=3.5 in]{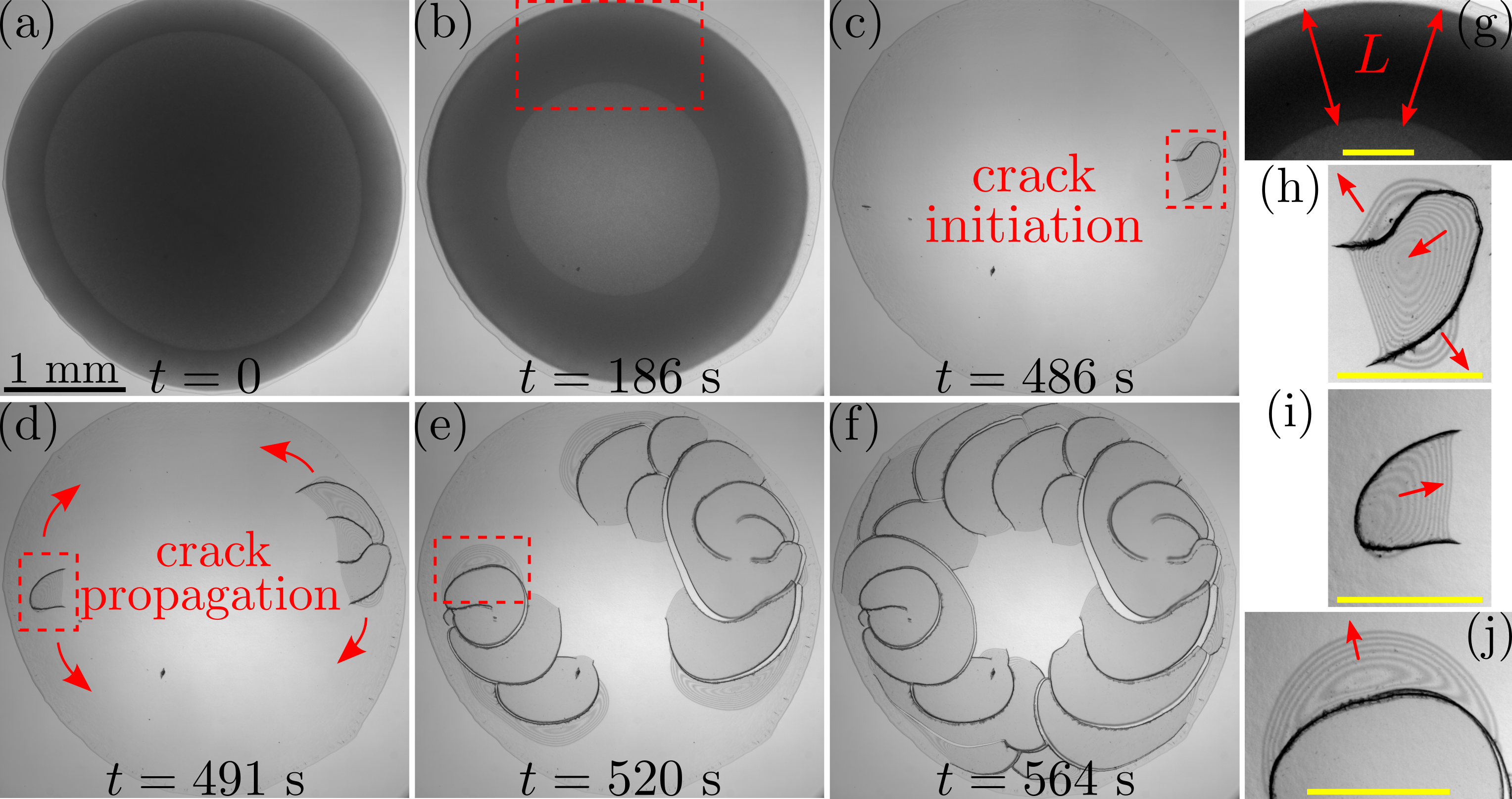}
\caption[]{Snapshots (a)-(f) show the characteristic stages during the formation of spiral-like cracks by drying a tumbler drop with $V_{i}=$ 2.5 $\mu$L and $\phi_{i}=14\%$. The drying process lasted about 10 min before the film completely dried. Image (g) shows the zoom-in view of the red boxed area in (b), where $L$ indicates the width of the compaction front. Images (h) and (i) show the firstly initiated crack in the boxed regions in (c) and (d), respectively. Image (j) shows the boxed area in (e), which displays the propagation of the delamination front along the circumferential direction as indicated by the arrow. The scale bars in (g)-(j) are 400 $\mu$m. Video S3 in ESI$^\dag$ shows the complete drying process corresponding to (a)-(f).
\label{Fig:crack_formation_tumbler}
} 
\end{center} 
\end{figure}

Figures\ \ref{Fig:crack_formation_tumbler}a-f show the snapshots of different stages during the formation of the spiral-like cracks by drying a drop of a tumbler suspension with $V_i=2.5$ $\mu$L and $\phi_i$ = 14 \%. The compaction front of the drying tumbler suspension was significantly wider than that of the drying swimmer suspension (Fig.\ \ref{Fig:cracks_swimmers}) and increased in width over time (comparing Videos S1 and S3 in ESI$^\dag$). As the accumulated stress reached the material strength, hairpin-shaped cracks facing the center of the drying film first appeared (boxed regions in Figs.\ \ref{Fig:crack_formation_tumbler}c and d), which simultaneously initiated film delamination between the two arms of the hairpins (interference fringes in Figs.\ \ref{Fig:crack_formation_tumbler}h and i). Upon the creation of new surfaces by cracking, the film subsequently delaminated perpendicularly to and outside the arms of the hairpins and propagated along the circumferential direction (Fig.\ \ref{Fig:crack_formation_tumbler}h). The delamination front showed an arc shape, which triggered the formation of a crack of the same shape at the front (Figs.\ \ref{Fig:crack_formation_tumbler}e and j). The crack in turn created a new surface for further delamination. This cycle of cracking and delamination repeated with time until the cracks initiated by different hairpins met, which ultimately resulted in the spiral-like crack pattern shown in Fig.\ \ref{Fig:crack_formation_tumbler}f. Thus, the spiral-like cracks of immotile bacteria stem from the dynamic interplay between film cracking and delamination.

\begin{figure*}[!]
\begin{center}
\includegraphics[width=1\textwidth]{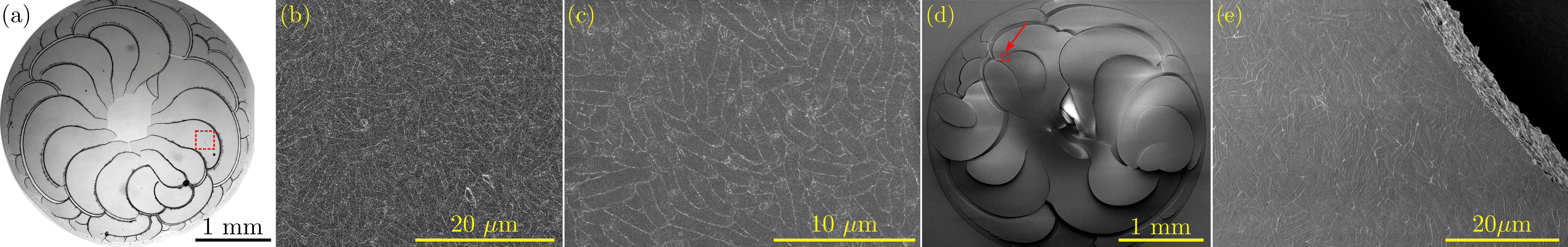}
\caption[]{(a) Bright-field image of a dried tumbler deposit with $V_i=2.5$ $\mu$L and $\phi_i$ = 20 \%. (b)-(c) SEM images of an area away from cracks enclosed by the red box in (a) with different magnifications. (d) SEM image of another dried tumbler deposit with $V_i=2.5$ $\mu$L and $\phi_i$ = 20 \%. (e) The zoom-in view of the area near a spiral-like crack enclosed by the red box as indicated by the red arrow in (d).
\label{Fig:crack_tumbler_SEM}
} 
\end{center} 
\end{figure*}

Due to the absence of the collective swimming, tumblers were isotropically distributed in a dried deposit as evidenced in Fig.\ \ref{Fig:crack_tumbler_SEM}, in contrast to the local order structure in the dried deposit of wild-type swimmers (Fig.\ \ref{Fig:SEM_swimmers}b). Figure\ \ref{schematic_setup}f further shows the height profile of the dried deposit of tumblers. By comparing Figs.~\ref{schematic_setup}c and f, one can see that, near the inner drop edge, the degree of delamination of the deposit of swimmers was much stronger than that of tumblers. The observation indicates that the dried deposit of active swimmers was thicker near the edge than that of immotile tumblers, as the bending moment responsible for film delamination is proportional to the film thickness\cite{giorgiutti2015dynamic}. Thus, from the conservation of the number of bacteria, the thickness of the deposit of tumblers should be more uniform than that of swimmers. The conclusion is further supported by the direct observation that the extent of the dried deposit of tumblers was larger than that of swimmers (comparing Figs.~\ref{schematic_setup}b and e). The more uniform deposit of tumblers likely arose due to the underlying uniform amorphous structure of bacteria with a relatively low packing density. In contrast, swimmers exhibited a local order structure of a high packing density near the pinned contact line, while maintained an amorphous structure of a low packing density near the center.

Our systematic control experiments on the drying suspensions of mutant tumblers and dead wild-type swimmers demonstrate that the local order structure driven by the collective swimming of wild-type bacteria plays an important role in the formation of circular cracks. More broadly, our experiments show that the crack patterns in the consolidating films of microorganisms can be tuned by manipulating the swimming behaviors of microorganisms.

\section{Discussion and conclusions}
It is interesting to compare the similarities and differences between crack patterns of drying bacterial suspensions and drying suspensions of colloidal ellipsoids. First, counterintuitively, the crack pattern of immotile bacteria is different from that of colloidal ellipsoids. In drying suspensions of colloidal ellipsoids, driven by the hydrodynamic torque of capillary flow, the ellipsoids align with their major axes parallel to the pinned contact line, which leads to a nematic order structure near the contact line. As cracks propagate preferably along the order direction, this order structure gives rise to circular cracks in the dried deposit \cite{dugyala2016role}. In contrast, immotile bacteria show an amorphous structure in the dried deposit regardless of their relative locations to cracks (Fig.~\ref{Fig:crack_tumbler_SEM}). Such an amorphous structure does not support the order-induced crack propagation. Hence, the crack patterns of colloidal ellipsoids and immotile bacteria are qualitatively different. We hypothesize that the amorphous structure of bacterial films arises because of the presence of randomly orientated flagella and pili on the surface of the body of immotile bacteria, which prevent immotile bacteria from packing orderly. 

Second, although circular cracks were observed in both drying suspensions of wild-type swimmers and suspensions of colloidal ellipsoids, the underlying mechanisms are quite different. As discussed in the previous paragraph, the circular cracks of colloidal ellipsoids form due to the nematic order of particles near the pinned contact line \cite{dugyala2016role}. While swimming bacteria do form a local order structure in the form of swarming vortices, the structure does not yield a large-scale system-wide nematic order along the contact line (Fig.~\ref{Fig:SEM_swimmers}). Instead, the large-scale circular cracks of drying bacterial films form due to the combined effect of the tensile stress along the radial direction $\sigma_{rr}$ and the compressive stress along the azimuthal direction $\sigma_{\theta\theta}$, as shown by our simple model.

In conclusion, we investigated the mechanical instabilities of evaporation-driven consolidating dense bacterial suspensions, a model system of active matter. Circular cracks were observed in the consolidating films of wild-type swimming \textit{E. coli}, which were followed by film delamination along the radial direction toward the center of the drying drop. We showed that the circular cracks are initiated by the local order structure of bacteria, which arises from the collective swimming of bacteria in the drying suspensions. The propagation of the circular cracks at large scales is then determined by the tensile nature of the radial drying stress, which were calculated within the framework of elastic fracture mechanics and poroelastic theory. The tensile radial stress in combination with the humidity gradient across the film thickness also leads to the delamination of the consolidating films upon the creation of the free surface by cracking. Moreover, we also observed spiral-like cracks in the dried deposits of immotile bacteria. Such an intriguing pattern stems from the complex interplay between cracking, delamination, film geometry and anisotropic drying stresses. The detailed mechanism of this complicated process is an open question for future research. It should be noted that our simplified model does not capture all of the factors involved in the complex active system of swimming bacteria, particularly those related to biological properties such as quorum-sensing and the variations of bacterial shapes. We hope our findings and simple analyses in these particular experimental settings can serve as a basis for stimulating more experimental, numerical and theoretical works on drying active matter in the future.

Taken together, our results elucidate the critical role of the microscopic bacterial activity on the macroscopic mechanical instabilities and pattern formation of consolidating bacterial films. Practically, our study provides insights into diverse biological, environmental and industrial processes associated with drying bacterial suspensions, such as spreading of pathogens, biofilm formation, painting and coating of biological fluids.

\section*{Conflicts of interest}
There are no conflicts to declare.

\section*{Acknowledgements}
We thank Justin Burton and Xudong Liang for fruitful discussions, and Greg Haugstad, Shashank Kamdar, Samantha Porter and Yiming Qiao for the assistance in the experiment. We acknowledge the AISOS at University of Minnesota for the access of Nanoveal Profilometer. This work was supported by NSF CBET-2028652.

%%%END OF MAIN TEXT%%%


%
\setcounter{equation}{0}

\begin{appendices}
\renewcommand{\theequation}{A\arabic{equation}}
\section{Derivation of the in-plane stresses $\sigma_{rr}$ and $\sigma_{\theta\theta}$}
For linearly poroelastic solids, the stress and strain are related by the Biot constitutive equation \cite{biot1941general,wang2017theory}:
\begin{eqnarray}
\epsilon_{i j} & = & \frac{1+\nu}{E} \sigma_{i j}-\frac{\nu}{E} \sigma_{k k} \delta_{i j}+\frac{\alpha (P_{pore}-P_{atm})}{3 K} \delta_{i j} \nonumber \\
& = &\frac{1+\nu}{E} \sigma_{i j}-\frac{\nu}{E} \sigma_{k k} \delta_{i j}-\frac{\alpha P^\star}{3 K} \delta_{i j}, 
\label{Biot_constitutive1}
\end{eqnarray}
where $\alpha \approx 1$ is the Biot‐Willis coefficient \cite{wang2017theory}, $\nu$ and $K=E/3(1-2\nu)$ are Poisson's ratio and bulk modulus of the film, respectively. Here, $E$ is the Young's modulus of the film. $P^\star =P^\star(z,t)$ is given by Eq.~\ref{pore_pressure_difference} in the main text.  $(i, j)$ take the values of $(r, \theta, z)$ in a cylindrical coordinate system. Accordingly, $\sigma_{k k}=\sigma_{rr}+\sigma_{\theta \theta}+\sigma_{z z}$.

Summing up the three principle strains according to Eq.\ \ref{Biot_constitutive1} under the approximations $\epsilon_{zz}\gg \epsilon_{rr}+\epsilon_{\theta\theta}$ and $\sigma_{zz}\ll \sigma_{rr}+\sigma_{\theta\theta}$ explained in the main text leads to: 
\begin{equation}
\epsilon_{zz}=\frac{1}{3 K}\left( \sigma_{rr}+\sigma_{\theta \theta} \right)-\frac{P^\star}{K}.
\label{Biot_constitutive}
\end{equation}
In combination with $\epsilon_{zz}=-\frac{\nu}{E}(\sigma_{rr}+\sigma_{\theta \theta}) -\frac{P^\star}{3 K}$ from Eq.\ \ref{Biot_constitutive1}, we have:
\begin{equation}
\sigma _{\theta \theta} +\sigma_{rr}=\frac{2 P^\star}{3 K (\frac{1}{3 K}+\frac{\nu}{E})}=\frac{2P^\star(1-2\nu)}{1-\nu}.
\label{in-plane_stress}
\end{equation}
Plugging Eq.\ \ref{in-plane_stress} into Eq.~\ref{radial_azimuthal_stress} in the main text leads to:
\begin{equation}
r\frac{ \partial\sigma_{rr}}{\partial r} +2 \sigma_{rr}  =2P^\star \left(2+\frac{1}{\nu-1}\right).
\label{eq:sigmarr}
\end{equation}
Since $P^\star$ shows small variations across the film thickness, we calculate the film-thickness-averaged liquid pressure to remove the weak $z$ dependence:
\begin{align}
\left\langle P^\star \right\rangle& =\frac{1}{h}\int_{-h}^0 \left[ \frac{2 \eta V_E}{\kappa}\sqrt{\frac{E\kappa t}{\pi \eta}} \mathrm{e}^{\frac{-z^{2}}{ 4 E \kappa t/\eta}}-\frac{ \eta V_Ez}{\kappa} \operatorname{erfc}\left( \frac{-z}{2 \sqrt{E \kappa t/\eta}}\right) \right] dz  \nonumber \\
&=  \frac{ V_E E t }{h}\left[1+\frac{h\mathrm{e}^{-\frac{h^2}{4ct}}}{\sqrt{\pi ct}}- \frac{h^2+2ct}{2ct} \operatorname{erfc} \left( \frac{h}{2}\sqrt{\frac{1}{ct}} \right) \right],
\label{eq:p_pore_average}
\end{align}
where $c \equiv E \kappa /\eta$.

The stress boundary condition at $r=R$ is $\sigma \cdot \hat{n}\mid_{r=R}=\sigma_{rr}\mid_{r=R}=P_{atm}-P_{pore}=P^\star$\cite{goehring2015desiccation}. Applying the condition and replacing $P^\star$ with $\left\langle P^\star \right\rangle$ (Eq.\ \ref{eq:p_pore_average}), Eq.\ \ref{eq:sigmarr} can be solved:
\begin{equation}
\sigma_{rr}= \left\langle P^\star \right\rangle \frac{r^2(2\nu-1)-R^2\nu}{r^2(\nu-1)}.
\label{sol:sigmarr}
\end{equation}
Plugging Eq.\ \ref{sol:sigmarr} into Eq.~\ref{Biot_constiutive}  in the main text yields:
\begin{equation}
\sigma_{\theta\theta} =  \left\langle P^\star \right\rangle \frac{r^2(2\nu-1)+R^2\nu}{r^2(\nu-1)}.
\label{sol:sigmatheta}
\end{equation}

\setcounter{figure}{0}
\setcounter{equation}{0}
\section{Critical stress for cracking $\sigma_c$ and for delamination $\sigma_d$}
\renewcommand{\theequation}{B\arabic{equation}}
\renewcommand{\thefigure}{B\arabic{figure}}

\begin{figure*}[h]
\begin{center}
\includegraphics[width=5.5 in]{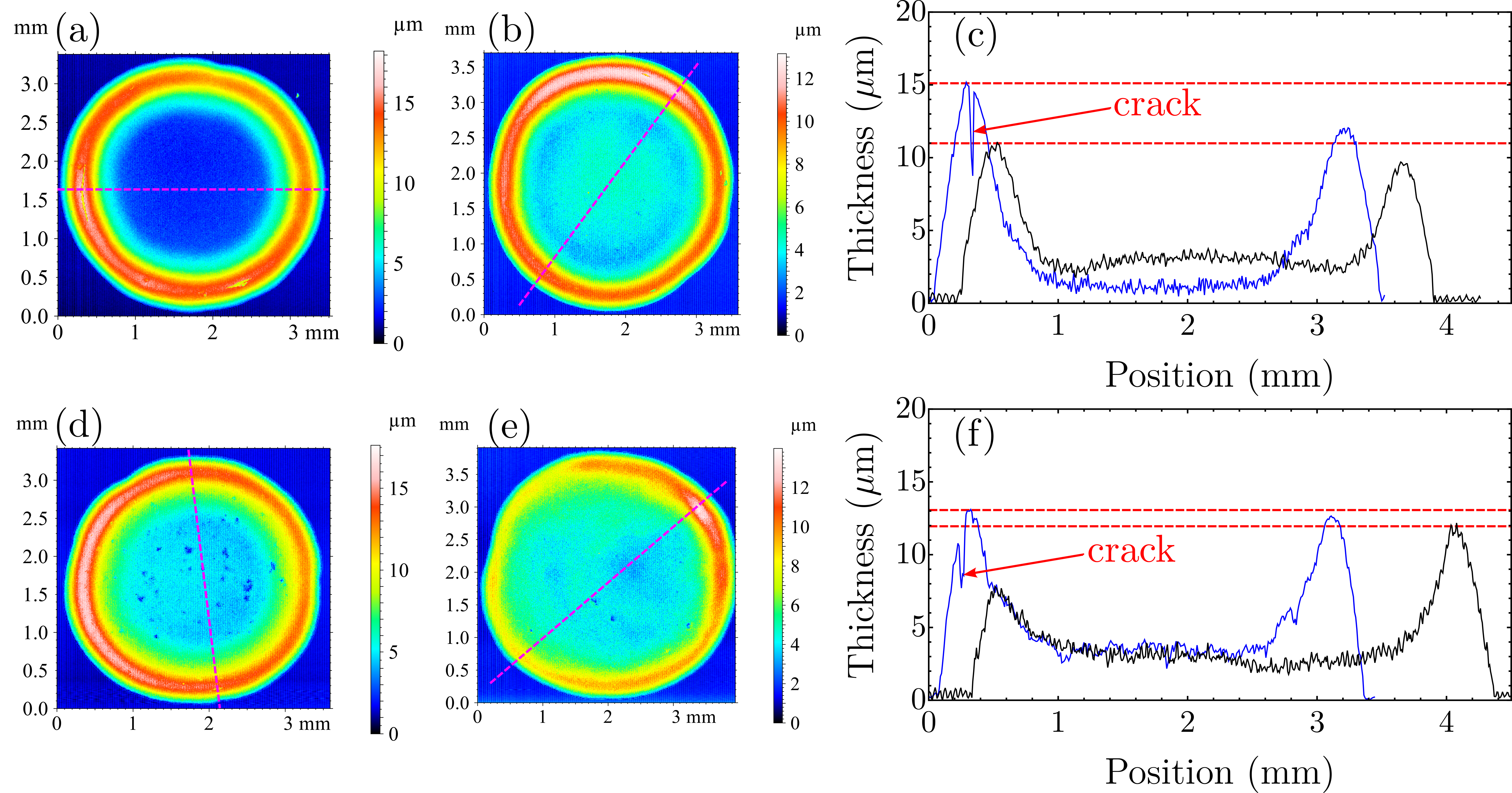}
\caption[]{Height profiles of two dried swimmer deposits (a and b) and two dried tumbler deposits (d and e) with two different film thicknesses. Cracks are present in the thicker films in (a) and (d), but not in the thinner films in (b) and (e). The blue and black curves in (c) and (f) represent the height profiles along the diameters marked by the dashed lines in (a)-(b) and (d)-(e), respectively. The sharp drops of the blue curves at $r \approx 0.35$ mm in (c) and at $r \approx 0.3$ mm in (f) indicate the presence of cracks. Before delamination at long times, the heights of drying drops are significantly smaller than those shown in Figs. ~\ref{schematic_setup}c and \ref{schematic_setup}f.}
\label{critical_film_thickness}
\end{center} 
\end{figure*}

An evaporation-driven consolidating film is prone to crack to release the excess stress once the accumulated stress is beyond the critical stress $\sigma_c$. The value of $\sigma_c$ can be determined by measuring the corresponding critical film thickness $h_c$ as \cite{tirumkudulu2005cracking,singh2007cracking}:
\begin{equation}
\frac{\sigma_{c} a}{2 \gamma}=0.1877\left(\frac{2 a}{h_{c}}\right)^{2 / 3}\left(\frac{G M_c \phi a}{2 \gamma}\right)^{1 / 3},
\label{Eq:critical_stress_film}
\end{equation}
where $G=\frac{E_b}{2(1+\nu_b)}$ is the shear modulus of bacteria, $E_b$ and $\nu_b$ are the Young's modulus and Poisson's ratio of the dehydrated bacteria, $a$ is the characteristic radius, $\gamma$ is the surface tension, $M_c$ is the coordination number, $\phi$ is the 3D packing fraction. To measure $h_c$, we prepared drops of bacterial suspensions with $V_i=2.5$ $\mu$L and various $\phi_i$, and deposited them on glass substrates for drying to obtain dried deposits of different thicknesses.

Figure\ \ref{critical_film_thickness} shows the measurements of the height profiles along the diameters of two dried deposits of swimmer drops (Figs.\ \ref{critical_film_thickness}a-c) and two dried deposits of tumbler drops (Figs.\ \ref{critical_film_thickness}d-f). In both cases, cracks were initiated in the dried deposits with the larger film thickness, but not in the deposits with the smaller film thickness. The critical film thickness for cracking, $h_c$, should lie between these two thicknesses. We simply took the average of the maximum values of the blue and black curves as indicated by the dashed lines in Figs.\ \ref{critical_film_thickness}c and f, which gave the estimate $h_c \approx 13$ $\mu$m for swimmers and $h_c \approx 12.5$ $\mu$m for tumblers. Plugging the typical values of $E_b =$ 300 MPa \cite{yao1999thickness}, $\nu_b =$ 0.2 \cite{yao1999thickness} $a = 1$ $\mu$m, $\gamma =72$ mN/m, $M_c=6$, and $\phi \approx 0.7$ for swimmers or $\phi \approx 0.6$ for tumblers (Appendix C) into Eq.\ \ref{Eq:critical_stress_film}, we have the critical stress $\sigma_c \sim 10^5$ Pa for both swimmers and tumblers.

The  critical stress for film delamination $\sigma_d$ can be estimated using \cite{meng2020cracking}:
\begin{equation}
\sigma_{d}=\sqrt{\frac{2 E \Gamma_{d}}{h_d}},
\label{eq:delamination_stress}
\end{equation}
where $\Gamma_{d}$ is the adhesion strength of the film, $E$ and $h_d$ are the Young's modulus and critical delamination thickness of the film, respectively. Although it is challenging to directly measure the adhesion between an evaporation-driven consolidating bacteria film and a glass substrate, the adhesion strength for most biofilms is on the order of magnitude of 5 $\mathrm{mJ / m^2}$ \cite{yan2019mechanical,chen1998direct, zhang2011interaction,dupres2005nanoscale,liu2010direct,jiang2021searching}. Here, the Young's modulus of the dehydrated bacteria films can be estimated as $E\sim 100 $ MPa. In our experiments, the film delamination and cracking developed nearly simultaneously. Therefore, we simply took $h_d \approx h_c \approx 13$ $\mu$m. Plugging the values of $\Gamma_{d}$, $E$ and $h_d$ into Eq.\ \ref{eq:delamination_stress} yields $\sigma_{d} \sim 10^5$ Pa for swimmer films, which is of the same order of magnitude as $\sigma_c$. Following the same analysis, $\sigma_d$ of tumbler films is also the same order of magnitude at $\sim 10^5$ Pa.

\setcounter{equation}{0}
\section{Estimate of the 3D packing fraction of bacteria in dried deposits}
\renewcommand{\theequation}{C\arabic{equation}}

To estimate the 3D packing fraction of bacteria in a dried deposit, we first calculate the relation between the 3D packing fraction $\phi$ and the 2D area fraction $\phi_{2d}$. Assuming a cylindrical shape of a bacterial body with a length of $l_b$ and radius of $r_b$. The volume of bacterial body is $V_0=\pi r_b^2 l_b$ and the cross-sectional area of the body along its major axis is $A_0=2 r_b l_b$. Consider bacteria within a horizontal layer of length $K$ and width $W$ parallel to the substrate. We assume that all the bacteria are confined within the layer with their major axes aligned parallel to the substrate, a configuration agreeing reasonably well with our SEM images (Figs.~\ref{Fig:SEM_swimmers} and \ref{Fig:crack_tumbler_SEM}). The number of bacteria within the layer is $N$. The 2D area fraction within the layer is given by $\phi_{2d}=\frac{N A_0}{KW}=\frac{2Nr_bl_b}{KW}$, whereas the 3D volume fraction is $\phi=\frac{N V_0}{KW 2r_b}=\frac{N\pi r_b l_b}{2KW}$. Thus, $\phi=\frac{\pi}{4}\phi_{2d}$. We experimentally measured $\phi_{2d}$ for wild-type swimmers based on the SEM images (Fig.\ \ref{Fig:SEM_swimmers}b), which gives $\phi_{2d}\approx 0.96$. Consequently, $\phi \approx 0.75$. We note that the experimental $\phi$ is slightly larger than the theoretical maximum values of 3D random packing fraction of rigid cylinders ($\approx 0.7$) \cite{donev2004unusually,li2010maximum,liu2018evolutions}, which is likely due to the local order structure of bacteria and the approximation taken in our estimate. Similarly, we also measured the 2D area fraction of tumblers in the dried deposit, which gives $\phi_{2d} \approx 0.75$. Thus, the 3D packing fraction of tumblers is: $\phi=\frac{\pi}{4} \phi_{2d} \approx 0.59$, which is smaller than the 3D packing fraction of the wild-type swimmers as expected.

\end{appendices}

\balance

\renewcommand\refname{References}

%%%REFERENCES%%%
\bibliography{rsc} %You need to replace "rsc" on this line with the name of your .bib file
\bibliographystyle{rsc} %the RSC's .bst file

\end{document}